\documentclass[usenatbib,usegraphicx,useAMS]{mn2e}
\usepackage{graphicx}
\usepackage{amssymb,url,times}
\usepackage{color}
\usepackage{ulem}

\title[Fluctuating neutron star magnetosphere]{Fluctuating neutron star magnetosphere: braking indices of eight pulsars, frequency second derivatives of $222$ pulsars and $15$ magnetars}

\author[Ou, Tong, Kou \& Ding]
     {Z. W. Ou$^{1,2}$, H. Tong$^1$\thanks{Corresponding author: tonghao@xao.ac.cn}, F. F. Kou$^{1,2}$, and G. Q. Ding$^1$\\
     $1$ Xinjiang Astronomical Observatory, Chinese Academy of Sciences, Urumqi, Xinjiang 830011, China\\
     $2$ University of Chinese Academic of Sciences, 19A Yuquan Road, Beijing 100049, China}

\begin{document}

\date{2016.1 v2}
\pagerange{\pageref{firstpage}--\pageref{lastpage}} \pubyear{2015}
\maketitle

\label{firstpage}
\begin{abstract}
Eight pulsars have low braking indices, which challenge the magnetic dipole braking of pulsars. 
$222$ pulsars and $15$ magnetars have abnormal distribution of frequency second derivatives, 
which also make contradiction with classical understanding. 
How neutron star magnetospheric activities affect these two phenomena are investigated by using the wind
braking model of pulsars. It is based on the observational evidence that pulsar timing is correlated with emission and both aspects reflect the magnetospheric activities. Fluctuations are unavoidable for a physical neutron star magnetosphere. 
Young pulsars have meaningful braking indices, while old pulsars' and magnetars' fluctuation item dominates 
their frequency second derivatives. It can explain both the braking index and frequency second derivative of pulsars uniformly. 
The braking indices of eight pulsars are the combined effect of magnetic dipole radiation and particle wind. 
During the life time of a pulsar, its braking index will evolve from three to one. 
Pulsars with low braking index may put strong constraint on the particle acceleration process 
in the neutron star magnetosphere. The effect of pulsar death should be considered during the long term
rotational evolution of pulsars. An equation like the Langevin equation for Brownian motion was derived for pulsar spin-down. 
The fluctuation in the neutron star magnetosphere can be either periodic or random, 
which result in anomalous frequency second derivative and they have similar results. 
The magnetospheric activities of magnetars are always stronger than those of normal pulsars.
\end{abstract}


\section{Introduction}           
Pulsars are rotating magnetized neutron stars (Gold 1968). Up to now, more than two thousand pulsars have been discovered. 
Magnetars are a special kind of pulsars. They may have much higher magnetic field 
than normal pulsars. Magnetars are assumed to be powered by the decay of their strong magnetic field
(Duncan \& Thompson 1992; Kouveliotou et al. 1998). Timing observations of pulsars and magnetars can give us some 
information about their dipole magnetic field. 
Once a pulsar's period and period derivative are measured, its characteristic dipole magnetic field is usually 
calculated by assuming magnetic dipole braking (Tong 2016)
\begin{equation}
\label{magneticstrength}
 B_{\rm c} =6.4 \times 10^{19} (P \dot{P})^{1/2} \,\rm G.
\end{equation}
The rotational evolution of the pulsar in the magnetic dipole braking model is
\begin{equation}
\label{Edot}
-I\Omega\dot{\Omega}=\frac{2 \mu^{2} \Omega^{4}}{3 c^{3}}\sin^{2}\alpha,
\end{equation}
where $\mu=\frac{B R^{3}}{2}$ is the neutron star's magnetic dipole moment, 
$I=10^{45} \,\rm g \,\rm cm^{2}$ is the neutron star's moment of inertia, and $\alpha$
is the magnetic inclination angle which is the angle between the rotational
axis and the magnetic axis.
The spin-down of pulsars can be described by a power-law (Lyne, Pritchard \& Smith 1993)
\begin{equation}
\label{nudotpowerlaw}
\dot{\nu} \propto \nu^{n},
\end{equation}
where the $\nu$ and $\dot{\nu}$ are pulsars' spin frequency and its derivative
respectively, $n$ is the so-called braking index. Observationally, the braking index is 
defined by the pulsar's frequency second derivative (Lyne et al. 2015)
\begin{equation}
\label{defn}
n=\frac{\nu \ddot{\nu}}{\dot{\nu}^2}.
\end{equation}
Besides this gradual trend of spinning down, real pulsars and magnetars also show various 
kinds of timing irregularities. These include glitches (Yu et al. 2013; Dib, Kaspi \& Gavriil 2008)
and timing noise (e.g., frequency second derivatives and subsequent timing residuals, Hobbs, Lyne \& Kramer 2010; Dib \& Kaspi 2014).

The magnetic dipole braking assumption predicts a braking index $n=3$ (or equivalently a predicted 
$\ddot{\nu} = 3 \dot{\nu}^2/ \nu$). However, two groups of observational phenomena challenge this assumption.
Firstly, the observations of eight pulsars reveal that all their steady braking indices are less than $3$. 
They are PSR B1509$-$58
($n=2.839\pm0.003$; Livingstone et al. 2005a; Livingstone \& Kaspi 2011), PSR
J1119$-$6127 ($n=2.684\pm0.002$; Weltevrede, Johnston \& Espinoza
2011), PSR J1846$-$0258 ($n=2.65\pm0.01$, Livingstone et al. 2006),
PSR B0531$+$21 (the Crab pulsar, $n=2.51\pm0.01$; Lyne, Pritchard \& Smith 1993; Lyne et al. 2015),
PSR B0540$-$69 ($n=2.14\pm0.009$; Livingstone, Kaspi \& Gavriil 2005b; Ferdman, Archibald \& Kaspi 2015),
PSR J1833$-$1034 ($n=1.857\pm0.001$; Roy, Gupta \& Lewandowski
2012), PSR B0833$-$45 (the Vela pulsar, $n=1.4\pm0.2$; Lyne et al. 1996), PSR
J1734$-$3333 ($0.9\pm0.2$; Espinoza et al. 2011).
Secondly, timing observations of more than 300 pulsars found an anomalous distribution of $\ddot{\nu}$ (Hobbs et al. 2010). 
Furthermore, timing observations of 15 magnetars have also found anomalous $\ddot{\nu}$ measurement (see Table 4 and references therein). 

The above two groups of observations will result in two consequences. 
Firstly, there may be other effects contribute to pulsar
braking. Secondly, these effects may have fluctuations which is unavoidable for a physical system.
One of the potential models which
can explain pulsars and magnetars spin-down is the wind braking model (Xu \& Qiao 2001; Kou \& Tong 2015).
The rotational evolution of the Crab pulsar has already been studied in the wind braking model by considering
the particle density and the effect of pulsar death (Kou \& Tong 2015). However, there
are other seven pulsars with measured braking indices. They include the high magnetic field pulsars
(PSR J1119$-$6127, PSR J1846$-$0258 and PSR J1734$-$3333) and the low braking index pulsars
(the Vela pulsar and PSR J1734$-$3333), which have different timing characteristics from the Crab
pulsar. Therefore, it is necessary to study the rotational evolution of all these eight pulsars.

The second aspect comes from that for some pulsars their timing variations are correlated
with changes in the pulse profile (Lyne et al. 2010; Keith, Shannon \& Johnston 2013; Brook et al. 2015).
Both the variation of timing and pulse profile may be caused by changes in the pulsar's magnetosphere.
The switched spin-down state of intermittent pulsars (Kramer et
al. 2006; Camilo et al. 2012; Lorimer et al. 2012) are also correlated with their
magnetospheric activities.
For a real magnetosphere, there will always be some fluctuations. After considering this aspect,
the braking indices of young pulsars and $\ddot{\nu}$ of old pulsars and magnetars can be understood uniformly.
The fluctuations of the magnetosphere can account for the measured $\ddot{\nu}$ of pulsars and magnetars. 

The wind braking model and understandings of eight pulsar's braking indices are presented in Section 2. 
The fluctuating neutron star magnetosphere and modeling of $\ddot{\nu}$ of pulsars and magnetars are shown in Section 3. 
Discussions and conclusions are given in Section 4 and Section 5, respectively.

\section{Braking indices of eight pulsars in the wind braking model}

\subsection{The wind braking model of pulsars}

Pulsars are oblique rotators in general. Xu \& Qiao (2001) proposed that both
the magnetic dipole radiation and particle outflow contribute to the braking torque of pulsars. 
In this wind braking model, the magnetospheric rotational energy loss rate is (Xu \& Qiao 2001; Kou \& Tong 2015)
\begin{equation}
\label{energyloss}
\dot{E}=\frac{2 \mu^{2} \Omega^{4}}{3 c^{3}}\eta.
\end{equation}
The rotational evolution of the pulsar is
\begin{equation}
\label{Edotwind}
-I\Omega\dot{\Omega}= \dot{E} =\frac{2 \mu^{2} \Omega^{4}}{3 c^{3}}\eta,
\end{equation}
where $\eta$ is a dimensionless function: $\eta = \sin^2 \alpha + 3 \kappa \Delta \phi/{\Delta \Phi} \cos^2 \alpha$. 
It depends on the magnetic inclination angle $\alpha$, the particle number density $\kappa$ 
(in units of Goldreich-Julian density, Goldreich \& Julian 1969), the particle acceleration potential $\Delta \phi$.
$\Delta \Phi$ is the maximum acceleration potential. 
For the vacuum gap model (VG(CR), Ruderman \& Sutherland 1975), 
$\eta=\sin^2 \alpha + 4.96\times 10^{2} \kappa B_{12}^{-8/7} \Omega^{-15/7} \cos^2 \alpha$, 
where $B_{\rm 12}$ is the magnetic field strength in units of $10^{12} \,\rm G$.
Expressions of $\eta$ depend on the specific acceleration model (see Table 2 of Kou \& Tong 2015). 

The braking index in the pulsar wind braking model is 
\begin{equation}
\label{nwind}
n = 3 + \frac{\Omega}{\eta} \frac{d \eta}{d \Omega}.
\end{equation}
The wind braking model includes the magnetic dipole radiation term (proportional to $\sin^2\alpha$) and the particle wind term
(proportional to $\cos^2\alpha$). The magnetic dipole radiation will result in a braking index $n=3$, while the particle wind term corresponds to a braking index of $n \approx 1$. Therefore, a braking index of $1 \leq n \leq 3$ of pulsars are expected in the wind braking model (Xu \& Qiao 2001). As a pulsar evolves, its braking index will change from about $3$
to about $1$ which means that the pulsar evolves from magnetic dipole braking dominated case to the wind braking dominated case
(Kou \& Tong 2015). 
For the long term rotational evolution of pulsars, the effect of pulsar death should be taken into consideration 
(Young, Manchester \& Johnson 1999; Contopoulos \& Spitkovsky 2006). 
The radio emissions will stop when the pulsars' rotational period approaches to the death period.
The death period is defined as (Kou \& Tong 2015):
\begin{equation}
\label{Pdeath}
P_{\rm death}=2.8 {(\frac{B}{10^{12}})}^{1/2} {(\frac{V_{\rm gap}}{10^{12}})}^{-1/2} \,\rm s,
\end{equation}
where $V_{\rm gap}$ is the maximum acceleration potential drop in the open
field line religions. For old pulsars, the effect of pulsar death should be taken
into account. Based on a series of previous works on wind braking of pulsars (Xu \& Qiao 2001; Li et al. 2014; Kou 
\& Tong 2015; Kou, Ou \& Tong 2015), it is applied to all the eight pulsars with braking index measured. 

\subsection{The braking indices and long term rotational evolution of eight pulsars}

The magnetic field strength, magnetic inclination angle and particle density are
assumed to be constant during the calculations\footnote{The case of variable particle density is considered in Kou et al. (2015).}.  
The VG(CR) model is taken as an example to calculate the rotational evolution of the eight pulsars.
The observational information of the eight pulsars are summarized in Table \ref{eightinput}. 
$B_{\rm c}$ is the characteristic magnetic field. 
$\tau_{\rm c}$ is the characteristic age $\tau_{\rm c}=\frac{P}{2 \dot{P}}$. $\tau_{\rm obs}$ is the observational age of pulsars. $P_{\rm 0}$ is the assumed initial rotational period  (Noutsos et al 2013; Igoshev \& Popov 2013; Gullon et al. 2014).
$\alpha$ is the observed (or assumed) magnetic inclination angle. 
 
\begin{figure}
\centering
\includegraphics[width=0.45\textwidth]{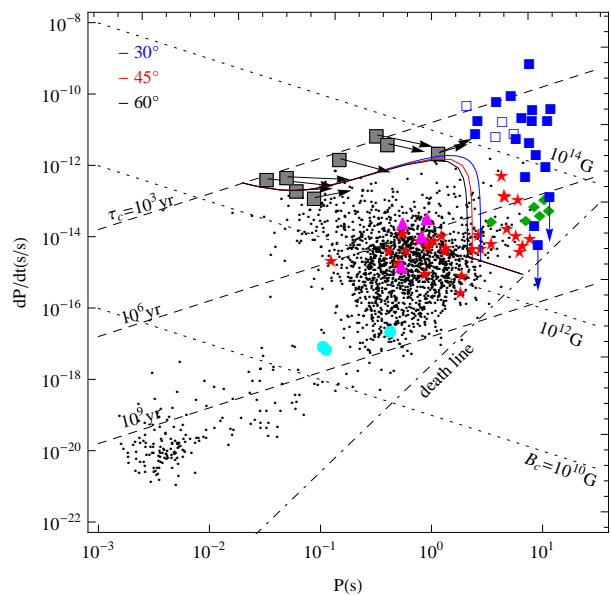}
\caption{Long term rotational evolution of PSR J1833$-$1034 with different inclination angle in the VG(CR) model. 
The blue, red, and black line represents inclination angle of $30^{\circ}$, $45^{\circ}$ and $60^{\circ}$, respectively.}
\label{J1833PPdotall}
\end{figure}

Given a pulsar's age, its initial rotational period $P_{\rm 0}$ can be calculated through equation (\ref{Edotwind}). However, the pulsars' age is poorly known. Therefore, an initial rotational period is assumed for each pulsar and its age in the wind braking model is calculated. Except for PSR J1734$-$3333, the other seven pulsars' initial rotational period are assumed to be $P_0 = 20\,\rm ms$. 
For PSR J1734$-$3333, considering its low braking index, an initial rotational period of $500\,\rm ms$ is adopted. 
PSR B0540$-$69 is taken as an example to show the different choice of initial rotational period. 
When this pulsar was given an initial rotational period of $10\,\rm ms$, $20\,\rm ms$ and $30\,\rm ms$,
the corresponding age are 2047 yr, 1736 yr and 1262 yr, respectively. There are only slight difference for their age. 
The observational age information of these eight pulsars are based on their associated supernova remnants
(Bocchino et al. 2005; Park et al. 2010; Kumar et al. 2012; Ho \& Anderson 2012). 
Most information about the magnetic inclination angle $\alpha$ are from fitting the light curve and spectra of $\gamma$ rays pulsars (Zhang \& Cheng 2000; Du et al. 2011; Du et al. 2012).
They include PSR B1509$-$58, the Crab pulsar, PSR B0540$-$69 and the Vela pulsar.
PSR J1846$-$0258, PSR J1833$-$1034 and PSR J1734$-$3333 lack inclination angle information, so an inclination angle of $45^{\circ}$ is used for these three pulsars. PSR J1833$-$1034 was taken as an example 
to test the difference choice of magnetic inclination angle. 
The corresponding results can be seen in Figure \ref{J1833PPdotall}. It shows that the evolutional traces in $P-\dot{P}$ 
diagram are not obviously different from each other, especially for young pulsars.  

The model parameters calculated from equation (\ref{Edotwind}) and (\ref{nwind}) are presented in Table \ref{eightoutput}. 
The magnetic fields of PSR J1119$-$6127 and PSR J1846$-$0258 can
reach to more than $10^{14} \,\rm G$ which are even higher than some of the magnetars. 
Comparing the magnetic field in the wind braking model to the characteristic magnetic field in Table \ref{eightinput},
PSR J1119$-$6127 and PSR J1846$-$0258 have similar values for these two factors. However, for PSR J1734$-$3333, the derived magnetic field
has obvious gap with the characteristic magnetic field. It may due to the presence of pulsar wind contribute extra braking torque to the pulsar. 
The Vela pulsar and PSR J1734$-$3333 have large age in the wind braking model among these eight pulsars, 
which may be the reason for their low values of braking index. 

\begin{figure}
\centering
\includegraphics[width=0.45\textwidth]{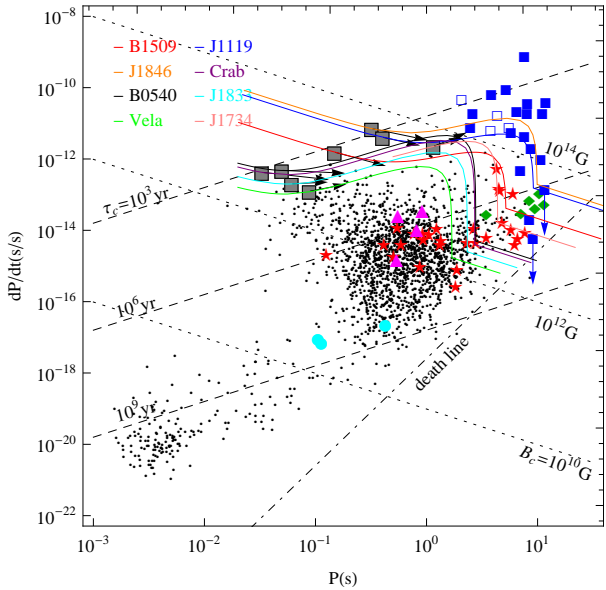}
\caption{Long term rotational evolution of the eight pulsars in the VG(CR) model. 
The $P$-$\dot{P}$ diagram of pulsars is updated from Figure 1 in Tong \& Wang (2014).}
\label{allsourcesPPdot}
\end{figure}

Based on the calculations above and parameters of Table \ref{eightinput} and Table \ref{eightoutput}, 
the long term rotational evolution of these eight pulsars are shown in Figure \ref{allsourcesPPdot}.
Except for PSR J1833$-$1034, the Vela pulsar and PSR J1734$-$3333, the other
five pulsars will first move to the lower right in the $P$-$\dot{P}$ diagram before they arrive the transition
point. The transition point is defined as when the pulsar braking index equals two. 
It means that there is a balance between the effect of magnetic dipole radiation and the effect of particle wind. 
Before they arrive at the transition point, the spin-down of pulsars
are dominated by magnetic dipole radiation and the braking indices values are between
$2$ and $3$. After that, pulsars will move to the upper right in the $P$-$\dot{P}$ diagram, 
and the spin-down of them are dominated by particle wind. During this process, the braking indices values
are between $1$ and $2$. PSR J1833$-$1034, the Vela pulsar and PSR J1734$-$3333 have
already been in this stage. However, they will not become magnetars.
They may pass through the magnetar domain, but all of them will move forward
lower right again considering the effect of pulsar death. 
Finally, after a pulsar stops radio emissions and the particle outflow ceases, 
the spin-down of pulsars is dominated by the magnetic dipole radiation again. 

\begin{table}
\begin{center}
\caption{Input parameters of eight pulsars. The columns are respectively: the pulsar's name, the observed braking index, 
the characteristic magnetic field, the characteristic age, an independent age observation, the assumed initial rotational period, 
and the observed (or assumed) magnetic inclination angle. References of braking indices are in Section 1.}
\label{eightinput}
\scriptsize
\begin{tabular}{lllllll}
\hline \hline
PSR name & $n$ & $B_{\rm c}$ & $\tau_{\rm c}$ & $\tau_{\rm obs}$ & $P_{\rm 0}$ & $\alpha$ \\
& & $(10^{12}\,\rm G)$ & $(\rm yr)$ & $(\rm yr)$ & $(\rm ms)$ & $(^{\circ})$ \\\hline

B1509$-$58 & 2.839(3) & 30 & 1555 & $1500^{\rm (1)}$ & 20 & $60^{\rm (9)}$ \\

J1119$-$6127 & 2.684(2) & 81 & 1607 & $7100^{\rm (2)}$ & 20 & $45^{\rm (10)}$ \\

J1846$-$0258 & 2.65(1) & 97 & 726 & $700^{\rm (3)}$ & 20 & 45 \\

B0531$+$21(Crab) & 2.51(1) & 7.5 & 1239 & $915^{\rm (4)}$ & 20 & $45^{\rm (11)}$ \\

B0540$-$69 & 2.14(1) & 9.9 & 1669 & $1000^{\rm (5)}$ & 20 & $50^{\rm (9)}$ \\

J1833$-$1034 & 1.857(1) & 7.1 & 4853 & $1000^{\rm (6)}$ & 20 & 45 \\

B0833$-$45(Vela) & 1.4(2) & 6.7 & 11303 & $9000^{\rm (7)}$ & 20 & $70^{\rm (12)}$ \\

J1734$-$3333 & 0.9(2) & 104 & 8128 & $8000^{\rm (8)}$ & 500 & 45 \\\hline
\end{tabular}
\flushleft
The observational age are from: (1) Gaensler et al. (1999); (2) Kumar, Safi-Harb \& Gonzalez (2012);
(3) Blanton \& Helfand (1996); (4) Lyne et al. (1993); (5) Park et al. (2010);
(6) Bocchino et al. (2005); (7) Page et al. (2009); (8) Ho \& Anderson (2012).\\

The observed magnetic inclination angle are from: 
(9) Zhang \& Cheng (2000); (10) Rookyard, Weltevrede \& Johnston (2015);
(11) Du, Qiao \& Wang (2012); (12) Du et al. (2011).
\end{center}
\end{table}

\begin{table}
\begin{center}
\caption{Calculated parameters of eight pulsars. The columns are respectively: the pulsar's name, the particle density,
the magnetic field strength (in units of $10^{12} \,\rm G$), the derived age in the wind braking model from an assumed initial period, 
the death period and the corresponding age when the pulsar stops radio emission.}
\label{eightoutput}
\scriptsize
\begin{tabular}{llllll}
\hline \hline
PSR name & $\kappa$ & $B_{\rm 12}$ & $\tau_{\rm w}$ & $P_{\rm d}$ & $\tau_{\rm d}$ \\
& $(10^3)$ & & $(\,\rm yr)$ & $(\,\rm s)$ & $(10^5 \,\rm yr)$ \\\hline

B1509$-$58 & 0.08 & 34 & 1588 & 5.17 & 1.97  \\

J1119$-$6127 & 0.025 & 107 & 1739 & 9.15 & 1.05 \\

J1846$-$0258 & 0.057 & 125 & 794 & 9.9 & 0.47 \\

B0531$+$21(Crab) & 0.57 & 9.4 & 852 & 2.71 & 0.79 \\

B0540$-$69 & 0.9 & 9.8 & 1764 & 2.77 & 0.59 \\

J1833$-$1034 & 0.41 & 6.9 & 6000 & 2.32 & 1.39 \\

B0833$-$45(Vela) & 1.75 & 3.6 & 19134 & 1.69 & 2.38 \\

J1734$-$3333 & 0.139 & 24.3 & 10444 & 4.37 & 0.93\\\hline
\end{tabular}
\end{center}
\end{table}

\subsection{Results for individual sources}

\begin{figure*}
\begin{center}
\centering
\includegraphics[width=0.9\columnwidth]{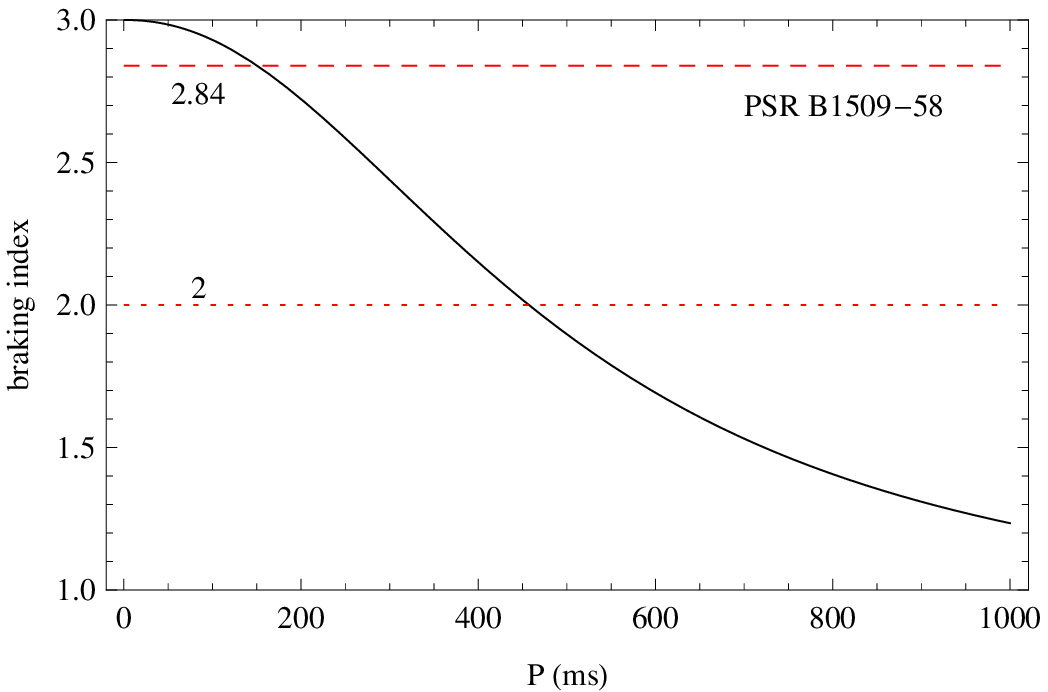}
\includegraphics[width=0.9\columnwidth]{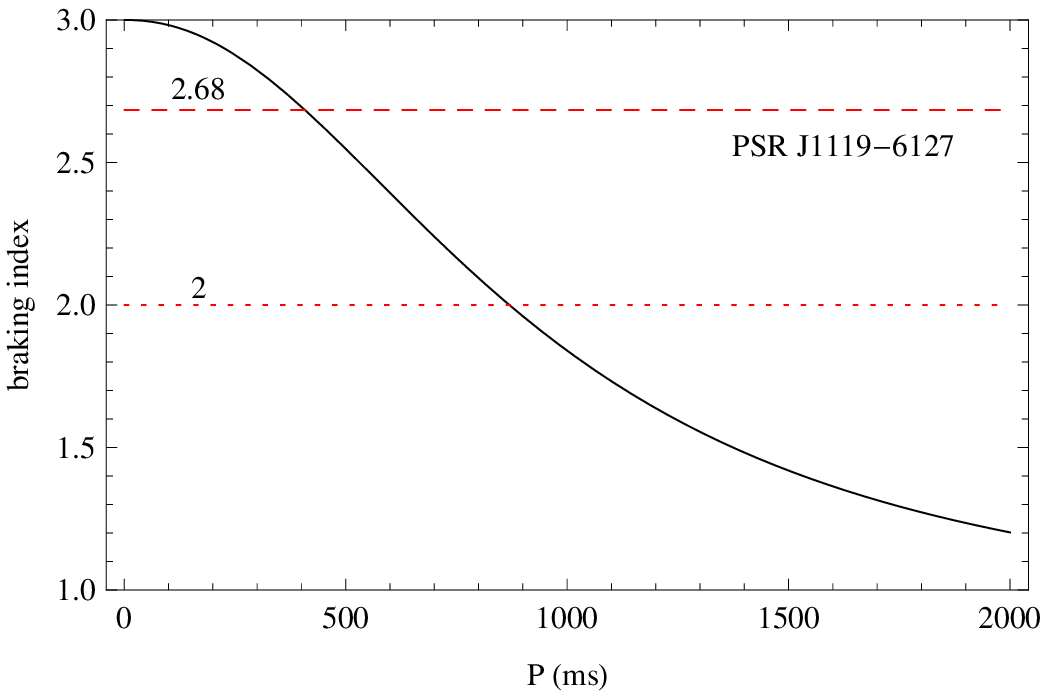}
\includegraphics[width=0.9\columnwidth]{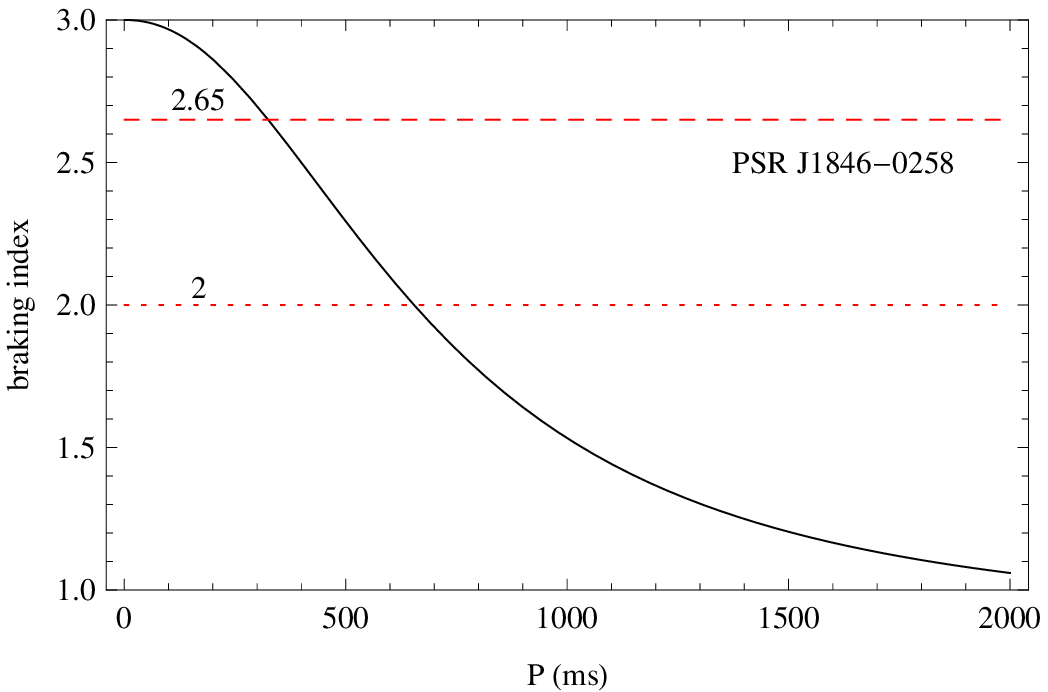}
\includegraphics[width=0.9\columnwidth]{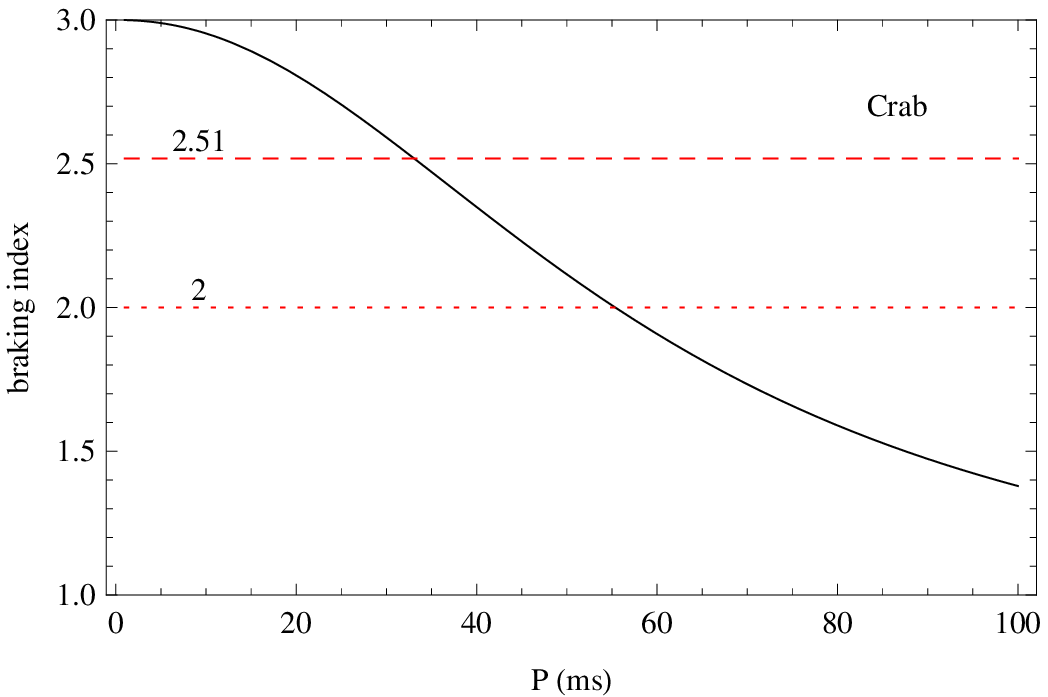}
\includegraphics[width=0.9\columnwidth]{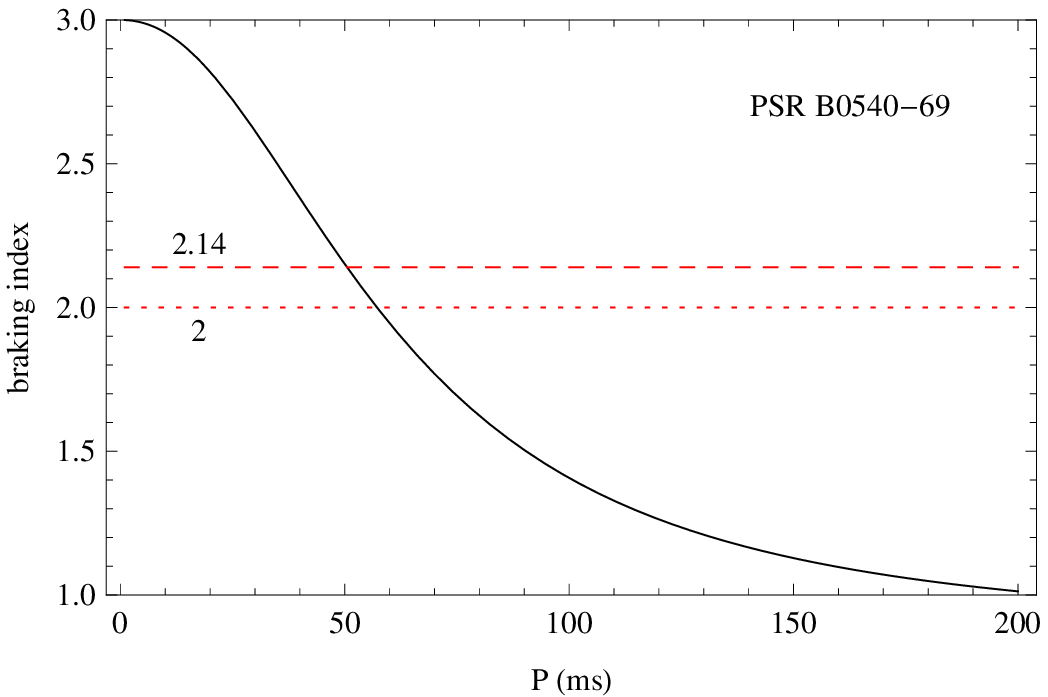}
\includegraphics[width=0.9\columnwidth]{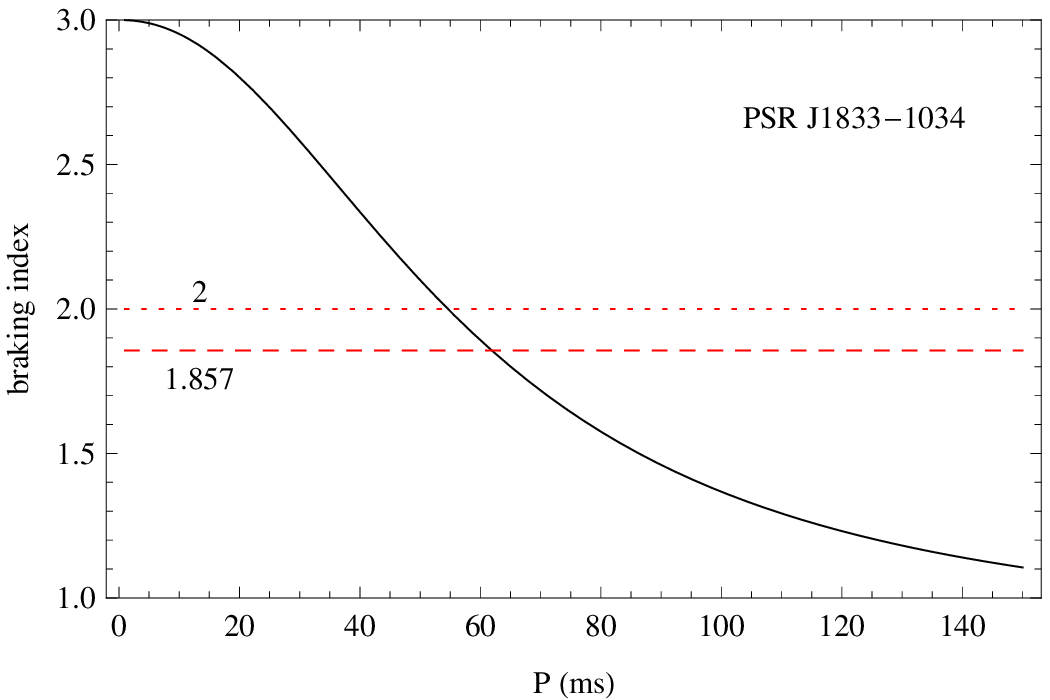}
\includegraphics[width=0.9\columnwidth]{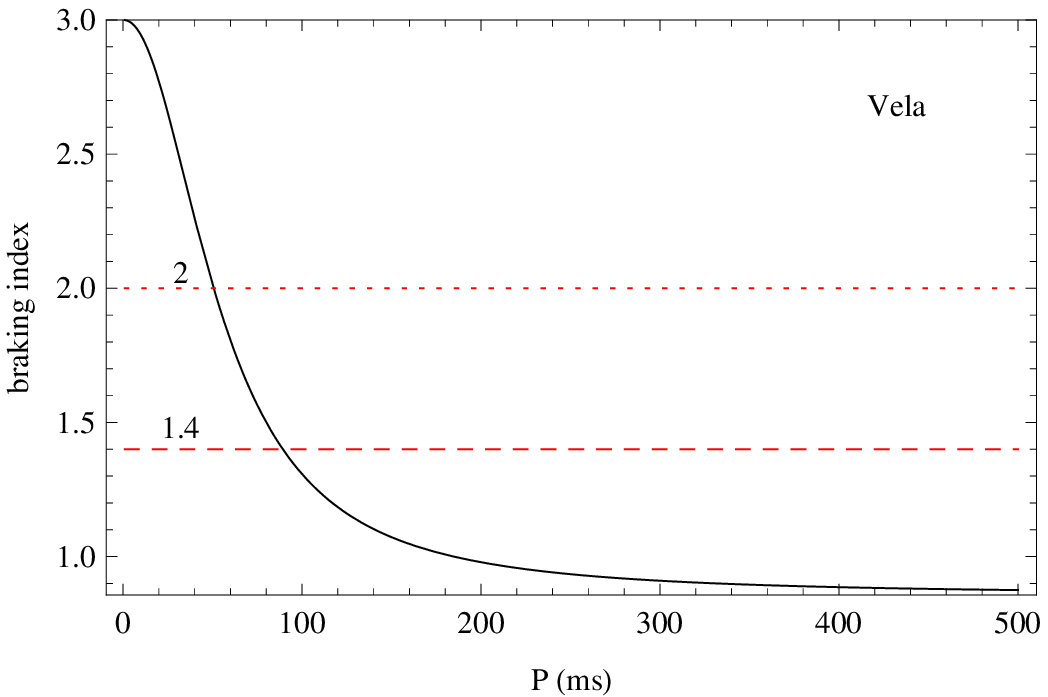}
\includegraphics[width=0.9\columnwidth]{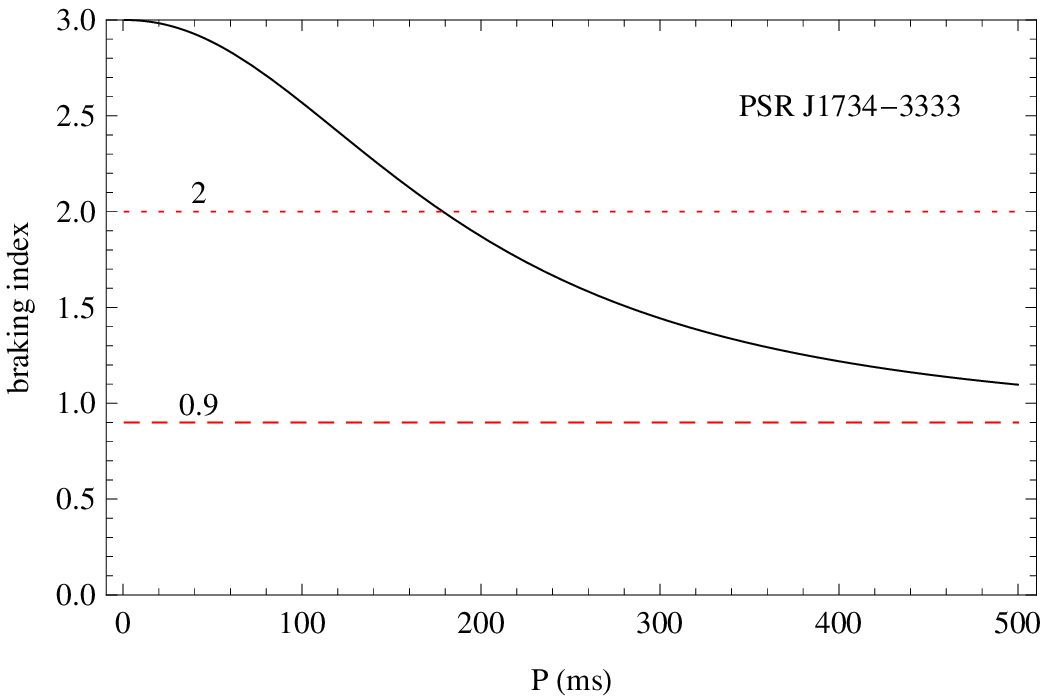}
\caption{Pulsar braking index as a function of their rotational period in the VG(CR) model. The dashed lines are observational value
of braking index and the dotted line is the transition value of $n=2$.}
\label{Eightgnp}
\end{center}
\end{figure*}

Figure \ref{Eightgnp} shows the braking indices of these eight pulsars as a function of their rotational period in the VG (CR) model.
The braking index will decrease from $3$ to $1$ as the rotational period increases. When $n=3$, the pulsar spin-down is dominated by its 
magnetic dipole radiation. With its evolution, the particle wind becomes stronger and stronger. A braking index of $n=2$ can be used to indicate
the moment when the magnetic dipole radiation and particle wind balances. After that, the particle wind continues to be
stronger until $n \approx1$, when the pulsar spin-down is dominated by the particle wind. 

\subsubsection{PSR B1509$-$58}

\begin{figure}
\centering
\includegraphics[width=0.45\textwidth]{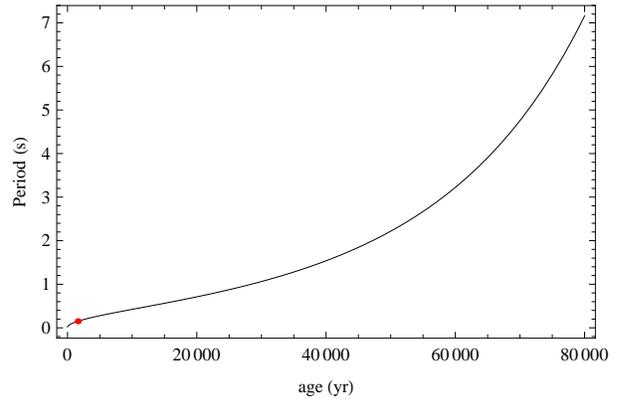}
\caption{Rotational period as a function of age for PSR B1509$-$58 in the VG(CR) model. 
The red point is its present period and age.}
\label{B1509}
\end{figure}

The braking index of PSR B1509$-$58 is $n=2.84$. It is near to the value of $n=3$, 
which is dominated by magnetic dipole radiation.
In the pulsar wind model, the spin-down of this pulsar is dominated by magnetic dipole
radiation while the particle wind still make some contribution. Therefore, the characteristic magnetic field
is similar to the derived magnetic field. 
And the characteristic age is similar to the derived age. Based on the three dimensional outer magnetosphere model
(Cheng, Ruderman \& Zhang 2000), the inclination angle of PSR B1509$-$58 is found to be about $60^{\circ}$ (Zhang \& Cheng 2000). 
Gaensler et al. (1999) concluded that PSR B1509$-$58 has an age of less than 1700 yr by estimating the distance and dispersion
measures of it. PSR B1509$-$58 was used as an example to calculate the short term evolution of period with age, 
which is shown in Figure \ref{B1509}. The pulsar evolves slowly in the early age which is dominated
by magnetic dipole radiation. Later, it evolves rapidly when the effect of particle wind dominates. 

\subsubsection{PSR J1119$-$6127}

The braking index of PSR J1119$-$6127 ($n=2.68$) is also near $3$.
Its spin-down is dominated by magnetic dipole radiation at present. The derived emission height
of PSR J1119$-$6127 is about 500 km, suggesting $\alpha \sim 17^{\circ} - 30^{\circ}$
(Weltevrede et al. 2011). Later, Rookyard, Weltevrede \& Johnston (2015)
pointed out that the region of inclination angle between $48^{\circ}$ and $144^{\circ}$ 
should be exclude. Kumar et al. (2012) estimated its associated supernova ramnant G292.5$-$0.5 age 
between 4.2 kyr and 7.1 kyr at an assumed distance of 8.4 kpc.

\subsubsection{PSR J1846$-$0258}

PSR J1846$-$0258 has a variable braking index (Archibald et al. 2015a; Kou et al. 2015). As it is dominated by magnetic
dipole radiation, its characteristic magnetic field is comparable with the derived magnetic field. 
Although Wang, Takata \& Cheng (2013) gave $10^{\circ}$ as the inclination angle of PSR J1846$-$0258, 
this value will lead to an extraordinary high magnetic field ($B=5.12\times10^{14} \,\rm G$). 
Therefore, an inclination of $45^{\circ}$ is taken during the calculations.
The death period of PSR J1846$-$0258 is the largest among the eight pulsars. 

\subsubsection{PSR B0531$+$21 (The Crab pulsar)}

The Crab pulsar has been monitored for a long term (Lyne et al. 1993; Lyne et al. 2015). 
The timing parameters of Lyne et al. (1993) is valid for AD 1969. 
As the pulsar was born at AD 1054, its age was 916 yr at AD 1969. 
Du et al. (2012) took $45^{\circ}$ as the magnetic inclination angle to calculate the light
curves of the Crab pulsar. The factors here are slightly different from Kou \& Tong (2015).
Although the observation age is certainly 916 yr, $20 \,\rm ms$ are taken as its initial period
and the corresponding age is 852 yr. The particle density here is $0.57 \times 10^{3} \rho_{\rm GJ}$
while it is $1.0 \times 10^{3} \rho_{\rm GJ}$ in Kou \& Tong (2015). 
Because the particle density is sensitive to the magnetic inclination angle.

\subsubsection{PSR B0540$-$69}

Like PSR J1846$-$0258, PSR B0540$-$69 also has variable timing behaviour 
(change of spin-down rate, Marshall et al. 2015; Kou et al. 2015).
Like PSR B1509$-$58, Zhang \& Cheng (2000) calculated the light curves and spectra of PSR B0540$-$69
and found an inclination angle of $50^{\circ}$. Based on various methods such as the pulsar spin-down, the kinematics 
of the optical ejecta, and the overall dynamics of the ejecta evolutionary models, the
age of its associated supernova remnant 0540-69.3 has been estimated to be 700-1600 yr (Park et al. 2010).
PSR B0540$-$69 has similar parameters ($\tau_{\rm obs}$, $B_{\rm 12}$ and $P_{\rm d}$) with the Crab pulsar. 

\subsubsection{PSR J1833$-$1034}

There are no inclination angle information about PSR J1833$-$1034, so an inclination angle of $45^{\circ}$ is adopted
during the calculations. Its characteristic age is similar to the derived age.
However, by applying a model of interaction between the pulsar wind nebula, the remnant and the supernova
environment, Bocchino et al. (2005) argued that its associated supernova remnant G21.5$-$0.9
age may be $200$-$1000\,\rm yr$, which is different from our derived value.

\subsubsection{PSR B0833$-$45 (the Vela pulsar)}

\begin{figure}
\centering
\includegraphics[width=0.45\textwidth]{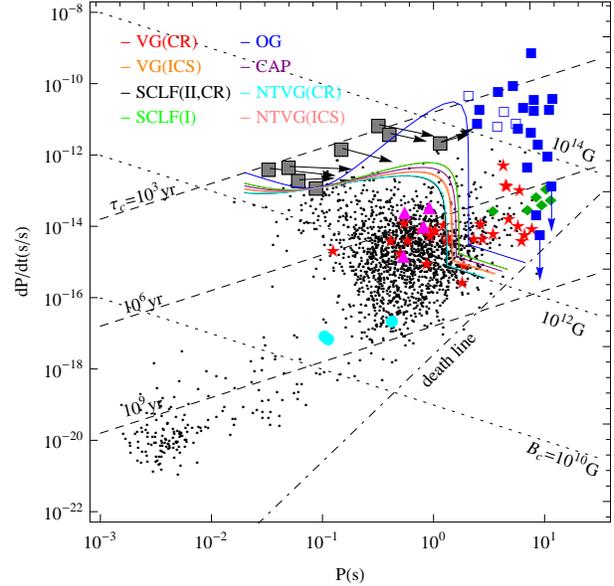}
\caption{Long term rotational evolution of the Vela pulsar in the different acceleration
models. The expression of $\eta$ can be seen in Table 2 of Kou \& Tong (2015). 
The top line is the OG model, 
while the bottom line is the SCLF(II,CR) and NTVG(CR) model. 
Since the power index of $\Omega$ in the VG(CR) model and SCLF(I)
are similar, their lines are coincident. For the same reason, 
the VG(ICS) model and the NTVG(ICS) model, 
the SCLF(II,CR) model and the NTVG(CR) model are similar, respectively.}
\label{VelaPPdotall}
\end{figure}

For pulsars with low braking index ($1 \leq n \leq 2$), there are two possible reasons. 
Firstly, when a pulsar was born it has already been dominated by the particle wind
($n \leq 2$). Secondly, the pulsar is very old so that the braking index evolves 
to the present low value. The Vela pulsar has the highest characteristic age among the eight pulsars.
Du et al. (2011) found that for the viewing angle of $64^{\circ}$, any magnetic
inclination angle between $60^{\circ}$ and $75^{\circ}$ in the annular gap model can
produce light curves with two sharp peaks and a large peak separation. 
Page et al. (2009) found that using $300\,\rm pc$ as the distance can give an age of 
the supernova remnant Vela about 5400-16000 yr as emphasized by Tsuruta et al. (2009).
Its long term rotational evolution in different acceleration models are calculated in Figure \ref{VelaPPdotall}. 
Each particle acceleration model has its own minimum value of braking index. They are listed in Table \ref{miniumindex}.
As for its low braking index ($n=1.4\pm0.2$), the Vela pulsar is a good source to constrain the particle acceleration models. 
The SCLF(II,ICS) model has already been ruled out in the case of the Vela pulsar.

\begin{table}
\begin{center}
\caption{The minimum braking index in each acceleration model. 
Except for the NTVG(CR) and NTVG(ICS) model, others are listed in Table 3 of Li et al. (2014). 
The SCLF(II,ICS) model can be ruled out for the Vela pulsar. 
As for PSR J1734$-$3333, only the VG(CR), SCLF(I) and OG models are meaningful.
VG stands for vacuum gap, SCLF stands for space charge limited flow, OG stands for outer gap,
CAP stands for constant acceleration potential, NTVG stands for near threshold vacuum gap. 
CR stands for curvature radiation, ICS stands for inverse Compton scattering.
See Li et al. (2014), Kou \& Tong (2015) and references therein for more information about each acceleration model.}
\label{miniumindex}
\scriptsize
\begin{tabular}{llllll}
 \hline\hline
 Model & $\rm VG(CR)$ & $\rm VG(ICS)$ & $\rm SCLF(II,CR)$ & $\rm SCLF(II,ICS)$ & $\rm SCLF(I)$ \\\hline
$n_{\rm min}$ & 0.86 & 1.1 & 1.3 & 2.4 & 0.86 \\\hline
Model & $\rm OG$ & $\rm CAP$ & $\rm NTVG(CR)$ & $\rm NTVG(ICS)$ & \\\hline
$n_{\rm min}$ & $-0.71$ & 1 & 1.24 & 1.12 & \\\hline
\end{tabular}
\end{center}
\end{table}

\subsubsection{PSR J1734$-$3333}

\begin{figure}
\centering
\includegraphics[width=0.45\textwidth]{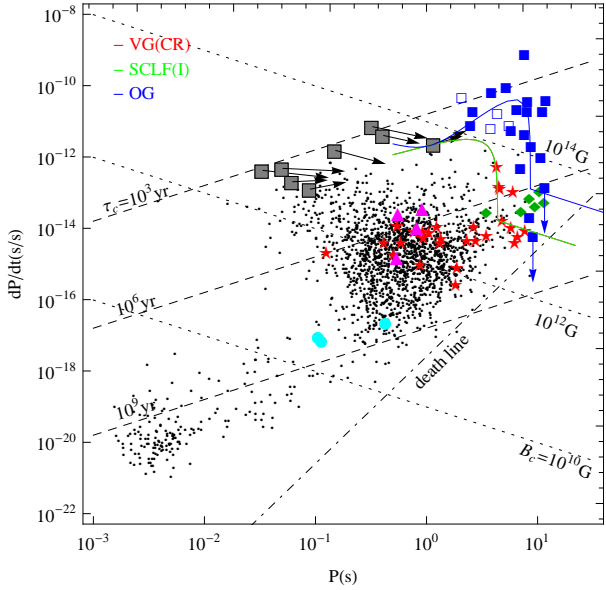}
\caption{Long term rotational evolution of PSR J1734$-$3333 in the different particle acceleration models. 
Only three particle acceleration models are drawn because the low braking index of PSR J1734$-$3333
is less than other models' minimum value of braking index. 
The red, green and blue line represents the VG(CR), SCLF(I) and OG model, respectively.
The red line and the green line are coincident.}
\label{J1734PPdot2all}
\end{figure}

PSR J1734$-$3333 lacks information about its inclination angle so an inclination angle of $45^{\circ}$ is used
during the calculations.
It has the highest characteristic magnetic field among the eight pulsars. 
Ho \& Anderson (2012) estimated its age by considering the supernova remnant size and expansion velocity to obtain an age about 2000 yr;
and considering the pulsar's distance away from the centre of the supernova remnant and pulsar space velocity, 
to obtain an age about 23000 yr. Like the Vela pulsar, PSR J1734$-$3333 has low braking index
value. Its death period is at the same level with its present period.
As a result, the death effect becomes a virtual factor for its rotational evolution.
Its long term rotational evolution in different acceleration models are calculated in Figure \ref{J1734PPdot2all}. 
Compared to the Vela pulsar, PSR J1734$-$3333 is a better source to constrain the particle acceleration models. 
It can be seen that only the VG(CR), SCLF(I) and OG models are valid for this pulsar.

\section{Frequency second derivatives of $222$ pulsars and $15$ magnetars considering magnetospheric fluctuations}

\subsection{Summary of $\ddot{\nu}$ observations of pulsars and magnetars}

Hobbs et al. (2010) presented the timing solutions for 366 pulsars, including 335 non-recycled pulsars\footnote{Recycled pulsars are defined as $P<0.1\,\rm s$ and $\dot{P}< 10^{-17}$ (Hobbs et al. 2010).}. From Table 1 in Hobbs et al. (2010), 
222 non-recycled pulsars have significant $\ddot{\nu}$ measurement\footnote{By saying significant, the criterion $|\ddot{\nu}|\ge 5\sigma_{\ddot{\nu}}$ is chosen.}. Among them,  117 have positive $\ddot{\nu}$, 105 have negative 
$\ddot{\nu}$. The remaining 113 non-recycled pulsars have no significant $\ddot{\nu}$ measurement. 
Therefore, there are roughly equal number of pulsars with positive, negative and no significant $\ddot{\nu}$. 
The $\ddot{\nu}$ measurement of magnetars lies in various individual timing papers. They are collected in Table 4. 
Up to now, a total of 15 magnetars have $\ddot{\nu}$ measurement.
Out of these 15 magnetars, 5 have positive $\ddot{\nu}$, 10 have negative $\ddot{\nu}$. 
7 confirmed magnetars\footnote{They are the three low-$\dot{P}$ magnetars: SGR 0418+5729, 3XMM J185246.6+003317, CXO J164710.2$-$455216; SGR 0526$-$66 (in LMC), CXOU J010043.1$-$721134 (in SMC), SGR 1627$-$41, and CXOU J171405.7$-$381031.}
have no $\ddot{\nu}$ reported 
(see the McGill online catalog for updates\footnote{http://www.physics.mcgill.ca/$\sim$pulsar/magnetar/main.html}, Olausen \& Kaspi 2014). 

\begin{figure}
\centering
\includegraphics[width=0.45\textwidth]{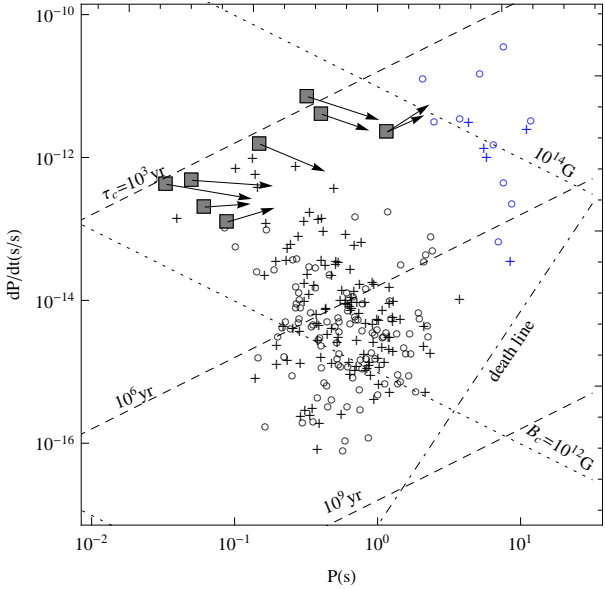}
\caption{Distribution of pulsars and magnetars with $\ddot{\nu}$ measured on the period-period derivative diagram. 
Black `+' means  pulsars with positive $\ddot{\nu}$ and black `$\circ$' for negative $\ddot{\nu}$. 
Blue `+' and `$\circ$' are for magnetars. Gray squares are pulsars with braking index measured. The arrow
marks the evolution direction of the pulsar (see Kou \& Tong 2015, figure 6 there for more explanations). }
\label{figPPdot}
\end{figure}

\begin{figure}
\centering
\includegraphics[width=0.45\textwidth]{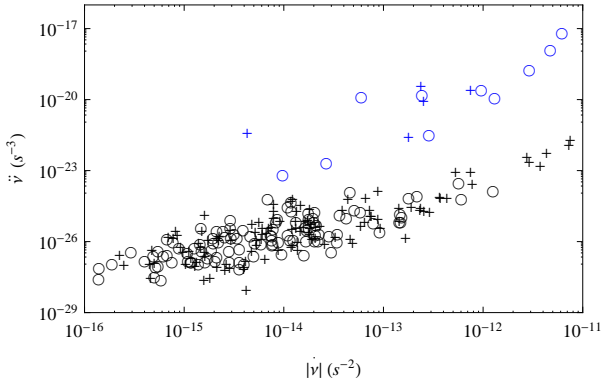}
\caption{$\ddot{\nu}$ versus $\dot{\nu}$ for pulsars and magnetars. Black `+' stands for pulsars with postive $\ddot{\nu}$
and black `$\circ$' for negative $\ddot{\nu}$ (from Table 1 in Hobbs et al. 2010). Blue `+' and `$\circ$' are for magnetars
(from Table 1).}
\label{fignu2dotobs}
\end{figure}

For pulsars and magnetars with $\ddot{\nu}$ measured, their distribution on the $P$-$\dot{P}$ diagram 
is shown in Figure \ref{figPPdot}.  It can be seen that only for young pulsars, their $\ddot{\nu}$ give a meaningful braking index. 
For old pulsars, their $\ddot{\nu}$ can be both positive and negative.  The $\ddot{\nu}$ as a function of $\dot{\nu}$ for  pulsars and magnetars are shown in Figure \ref{fignu2dotobs}. The observations show that: (1) The $\ddot{\nu}$ can be both positive and negative. It may reflect the timing noise of pulsars and magnetars, not the braking index. (2) The $\ddot{\nu}$ increase with $|\dot{\nu}|$, as had been shown by previous works (Hobbs et al. 2010). This correlation explains other correlations, e.g., correlation between $\ddot{\nu}$ and characteristic age etc. (3) Magnetars have a higher level of timing noise than normal pulsars (Woods et al. 2002). This is not unexpected since magnetars have more magnetospheric activities than normal pulsars (Mereghetti, Pons \& Melatos 2015). (4) The $\ddot{\nu}$ can change sign for the same source during different time span of observations (Hobbs et al. 2010; Dib \& Kaspi 2014). 

\begin{table}
\scriptsize
  \begin{center}
  \caption{Timing parameters of 15 magnetars. The columns are respectively: 
  the source name, the pulse frequency, the frequency derivative, the frequency second derivative, 
  and the references. The last digit uncertainties of $\ddot{\nu}$ are included, 
  usually 1$\sigma$ TEMPO-reported uncertainties.}
  \label{table_nu2dot}
  \begin{tabular}{lllll}
    \hline\hline
    source name & $\nu$ & $\dot{\nu}$ & $\ddot{\nu}$ & Refs.\\
     & $(\rm s^{-1})$ & $(\rm s^{-2})$ & $(\rm s^{-3})$ & \\
    \hline
    4U 0142+61 & 0.115 & $-2.679\times 10^{-14}$ & $-2.0(2)\times 10^{-23}$ & a\\
    
    SGR 0501+4516 & 0.173 & $-1.789\times 10^{-13}$ & $3.016\times 10^{-22}$ & b \\
    
    1E 1048.1$-$5937 & 0.155 & $-2.43\times 10^{-13}$ & $-1.62(8)\times 10^{-20}$ & c\\

    1E 1547.0$-$5408 & 0.483 & $-6.19\times 10^{-12}$ & $-6.69(7)\times 10^{-18}$ & d\\
    
    PSR J1622$-$4950 & 0.231 & $-7.5\times 10^{-13}$ & $2.9\times 10^{-20}$ & e\\
    
    1RXS J170849.0$-$400910 & 0.091 & $-2.38\times 10^{-13}$ & $400(50)\times 10^{-22}$ & f\\
    
    SGR J1745$-$2900 & 0.266 & $-9.60\times 10^{-13}$ & $-2.6(1)\times 10^{-20}$ & g \\
        
    SGR 1806$-$20 & 0.132 & $-4.73\times 10^{-12}$ & $-1.3(4)\times 10^{-18} $ & h \\
    
    XTE J1810$-$197 & 0.18 & $-2.53\times 10^{-13}$ & $9.40(6)\times 10^{-21}$ & i\\
    
    Swift J1822.3$-$1606 & 0.118 & $-4.3\times 10^{-15}$ & $4.4(6)\times 10^{-22}$ & j\\
    
    SGR 1833$-$0832 & 0.132 & $-6.0\times 10^{-14}$ & $-1.3(2)\times 10^{-20}$ & k \\
    
    Swift J1834.9-0846 & 0.403 & $-1.308\times 10^{-12}$ & $-1.2(3)\times 10^{-20}$ & l \\
   
     1E 1841$-$045 & 0.085 & $-2.866\times 10^{-13}$ & $-3.2(4)\times 10^{-22}$ & m\\
    
    SGR 1900+14 & 0.193 & $-2.913\times 10^{-12}$ & $-1.72(3)\times 10^{-19}$ & n\\    
    
    1E 2259+586 & 0.143 & $-0.973\times 10^{-14}$ & $-6.5(4)\times 10^{-24}$ & o\\       
    \hline
  \end{tabular}
\flushleft
\textbf{References:}

a: From Dib \& Kaspi (2014), Table 6 there. The ephemeris with the largest number of TOAs are selected. 

b: From Camero et al. (2014), Table 2 there. There frequency second derivative is converted from the period second 
derivative in that paper. Therefore, it includes no uncertainties. 

c: From Archibald et al. (2015b), Table 2 there. 

d: From Dib et al. (2012), Table 4 there.  

e: Estimated from figure 1 in Levin et al. (2012). $\dot{\nu}$ is approximately the median value between
MDJ 55100 to 55300. $\ddot{\nu}$ is estimated from $\dot{\nu}$ measurement during this time span. 

f: Also from Dib \& Kaspi (2014), Table 4 there. 

g: From Coti Zelati et al. (2015), Table 2 there, solution A. See Tong (2015) for alternative explanations. 

h: From Woods et al. (2007), Table 3 there. 

i: From Camilo et al. (2007), Table 1 and figure 3 there. 

j: From Scholz et al. (2012), Table 2 there. Timing solution 3 is selected, since it is the best guess by the 
authors. See Tong \& Xu (2013), Scholz, Kaspi \& Cumming (2014) for alternatives. 

k: From Esposito et al. (2011), Table 3 there. 

l: From Esposito et al. (2013), Table 2 there.

m: Also from Dib \& Kaspi (2014), Table 3 there. 

n: From Woods et al. (2002), Table 2 there.

o: Also from Dib \& Kaspi (2014), Table 5 there.

\end{center}
\end{table}

\subsection{Fluctuating neutron star magnetosphere}

The correlation between $\dot{\nu}$ variations and radio emissions indicates that 
the magnetospheric processes are responsible for both the change of torque on the neutron star and 
the radio emissions (Lyne et al. 2010; Keith et al. 2013; Brook et al. 2015). If the variation 
amplitude in $\dot{\nu}$ is 
less than $1\%$, then this change of torque may remain undetected by current pulsar timing and it may result in some kind of 
timing noise. Denote the time averaged rotational energy loss rate in the neutron star magnetosphere as $\dot{E}_{\rm steady}$,
then 
\begin{equation}
- I \Omega \dot{\Omega} =\dot{E}_{\rm steady}.
\end{equation} 
The slow down of the neutron star may be written in a general form
\begin{equation}
\label{steady_powerlaw}
\dot{\Omega} =-k \Omega^n,
\end{equation}
where the minus sign on the right hand of the above equation means the neutron star is slowing down, and $k$ may also depend 
on $\Omega$ etc.  The second time derivative of the angular velocity is 
\begin{equation}
\ddot{\Omega} = \frac{\dot{\Omega}^2}{\Omega} [  n+ \frac{\Omega}{\dot{\Omega}} \frac{\rm d}{{\rm d} t} \log k ]. 
\end{equation}
And the corresponding steady state braking index will be 
\begin{equation}
n_{\rm steady} =  n+ \frac{\Omega}{\dot{\Omega}} \frac{\rm d}{{\rm d} t} \log k. 
\end{equation}
If $k$ depends on time only implicitly through $\Omega$ which is true in the case of wind braking: 
$n=3$ and $k \propto \eta(\Omega)$, then 
\begin{equation}
\label{nwindsteady}
n_{\rm steady} = 3+ \frac{\Omega}{\eta} \frac{{\rm d} \eta}{{\rm d} \Omega}. 
\end{equation}
For young pulsars, this expression is valid, and it is consistent with Equation (\ref{nwind}). 
In the above expression for the braking index, 
it is assumed that the second frequency derivative of a pulsar is not contaminated by the noise process. 
Then the braking index reflects the torque dependence on $\Omega$. 
However, in the real case, fluctuations in the magnetosphere are unavoidable. 
The fluctuation in the neutron star magnetosphere can be modelled via periodic function or random variable.  
These two approximations give the same result. 

\subsubsection{Periodic fluctuations}

\begin{figure}
\centering
\includegraphics[width=0.45\textwidth]{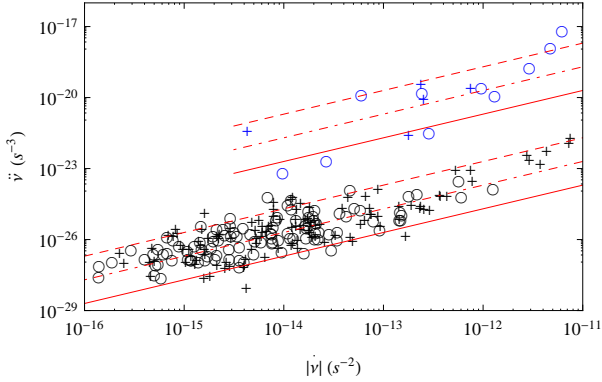}
\caption{Same as figure \ref{fignu2dotobs}, with the model calculations added (equation (\ref{nu2dot_periodic_case})). 
The time scale is chosen as $T=0.1\,\rm yr$ for both pulsars and magnetars. The lower three lines are 
$\delta=10^{-5},\ 10^{-6},\ 10^{-7}$ for the dashed, dot-dashed, and solid line, respectively.  
The upper three lines are $\delta=0.1,\ 0.01,\ 0.001$ for the dashed, dot-dashed, and solid line, respectively.}
\label{fignu2dotmodel}
\end{figure}

In the presence of some fluctuation in the neutron star magnetosphere, the magnetospheric torque may also fluctuate. 
Assuming periodic fluctuations, then the magnetospheric rotational energy loss rate is
\begin{equation}
\label{periodicEdot}
\dot{E} =\dot{E}_{\rm steady} (1+\delta \sin \frac{2\pi}{T} t),
\end{equation}
where $\delta$ is the amplitude of the fluctuation, and $T$ is the time scale. The corresponding slow down rate of the neutron star is 
\begin{equation}
\label{periodicomega}
\dot{\Omega} = - k \Omega^n (1+\delta \sin \frac{2\pi}{T} t).
\end{equation}
The angular velocity second derivative is 
\begin{eqnarray}
\label{periodicOmega2dot}
\ddot{\Omega} & =& \frac{\dot{\Omega}^2}{\Omega} [  n+ \frac{\Omega}{\dot{\Omega}} \frac{\rm d}{{\rm d} t} \log k
(1+\delta \sin \frac{2\pi}{T} t) ]\\
&=&  \frac{\dot{\Omega}^2}{\Omega} [  n_{\rm steady} + \frac{\Omega}{\dot{\Omega}} \frac{\rm d}{{\rm d} t} \delta \sin 
\frac{2\pi}{T}t ] \\ \label{Omega2dot_general}
&=& \frac{\dot{\Omega}^2}{\Omega} [  n_{\rm steady} - 4\pi \delta \frac{\tau_{
\rm c}}{T} \cos \frac{2\pi}{T}t ].
\end{eqnarray}
In the above deductions, a serial expansion of $\delta$ is made since fluctuation satisfies $\delta \ll 1$. 
From equation (\ref{Omega2dot_general}), for $4\pi \delta \frac{\tau_{\rm c}}{T} \gg 1$ or $\tau_{\rm c} \gg \frac{T}{4\pi \delta}$,
the $\ddot{\Omega}$ will be dominated by the fluctuation term. That is for old pulsars, 
their $\ddot{\Omega}$ are more likely to be dominated by the magnetospheric fluctuations. Since the fluctuation contribution contains a cosine term, this means that there should be roughly equal number of positive $\ddot{\Omega}$ and negative $\ddot{\Omega}$. 
At the same time, for some pulsars their $\ddot{\Omega}$ may be not very significant.  Only for young pulsars $\tau_{\rm c} \ll \frac{T}{4\pi \delta}$, their $\ddot{\Omega}$ may be dominated by the steady state case and the corresponding braking index is meaningful. For old pulsars, their corresponding braking indices have lost the original meaning. 
Therefore, in stead of employing the braking index, $\ddot{\Omega}$ will be used in the following. When the fluctuation
dominates, $\ddot{\Omega}$ is  
\begin{eqnarray}
\ddot{\Omega} &=& \frac{\dot{\Omega}^2}{\Omega}\times -4\pi \delta \frac{\tau_{\rm c}}{T} \cos \frac{2\pi}{T} t\\
&=& -2\pi \delta \frac{|\dot{\Omega}|}{T} \cos\frac{2\pi}{T} t. 
\end{eqnarray}
where $|\dot{\Omega}|$ is the absolute value of $\dot{\Omega}$. 
Observationally, the frequency is used in stead of angular velocity. 
The corresponding expression is 
\begin{equation}
\label{periodicnu2dot}
\ddot{\nu} = -2\pi \delta \frac{|\dot{\nu}|}{T} \cos\frac{2\pi}{T} t. 
\end{equation}
Besides $\delta$ and $T$, $\ddot{\nu}$ also depends on $t$. Then the $\ddot{\nu}$ may change sign during different time
span of observations, which is consistent with Hobbs et al. (2010).
Statistically, the amplitude of $\ddot{\nu}$ due to magnetospheric fluctuation with amplitude $\delta$ and time scale $T$ is 
\begin{equation}
\label{nu2dot_periodic_case}
|\ddot{\nu}| = 2\pi \delta \frac{|\dot{\nu}|}{T}. 
\end{equation}
The amplitude of $|\ddot{\nu}|$ is proportional to $|\dot{\nu}|$, which is also consistent with the observations (see Figure \ref{fignu2dotobs}). 
The same result can be obtained when assuming random fluctuation in the magnetosphere. 
Figure \ref{fignu2dotmodel} shows the model calculations along with the observational data. 
The amplitude of $\ddot{\nu}$ depends on the combination of $\delta/T$. The time scale is chosen 
as $T=0.1\,\rm yr$ for both pulsars and magnetars. In the real case, the time scale can vary from source to source. 
This choice has absorbed all the differences between different sources into the fluctuation amplitude. 
Magnetars have more magnetospheric activities are reflected in their larger values of fluctuation amplitude. 
It can be seen that most of the radio pulsars locate in the range of $\delta$ between $10^{-7}$ and $10^{-5}$
while most of the magnetars locate in the range of $\delta$ between $0.001$ and $0.1$. The magnetospheric
fluctuation of magnetars are more dramatic than that of normal pulsars.

\subsubsection{Random fluctuations}

The rotational energy loss rate due to the magnetospheric process may be made up of a steady component and a fluctuating component
\begin{equation}
\label{randomEdot}
\dot{E} = \dot{E}_{\rm steady} (1+\delta(t)),
\end{equation}
where $\delta(t)$ is a random variable. The slow-down of the neutron star is
\begin{equation}
\label{Omegadot_general}
\dot{\Omega} = -k \Omega^n (1+\delta(t)).
\label{randomomega}
\end{equation}
In general, the slow-down rate is determined by a dissipation term and  a fluctuation term. This is similar to the Langevin equation 
for Brownian motion (Pathria \& Beale 2011; Thorne \& Blandford 2016\footnote{We use some pre-final version of this book found on the web, e.g. http://pmaweb.caltech.edu/Courses/ph136/yr2012/}), except that the above equation is nonlinear in the general case, e.g. $n=3$ and $k$ also depends on $\Omega$. In order to linearize equation (\ref{Omegadot_general}), the angular velocity can be separated into a steady component and a fluctuating component
\begin{equation}
\label{perturbation}
\Omega(t) = \Omega_0(t) + \Omega_1(t),
\end{equation}
where $\Omega_0(t)$ is the angular velocity as function of time in the steady case, $\Omega_1(t)$ is a fluctuating term. According to the definition, $\Omega_0(t)$ satisfies: $\dot{\Omega}_0(t) = -k(\Omega_0) \Omega_0(t)^n$, where the dependence of $k$ on $\Omega$ is written explicitly. Substituting equation (\ref{perturbation}) into equation (\ref{Omegadot_general}), 
\begin{eqnarray}
\dot{\Omega}_{0} + \dot{\Omega}_{1} &=& -k(\Omega_0+\Omega_1) \cdot (\Omega_0+\Omega_1)^n \cdot (1+\delta(t))\\
&=& -k(\Omega_0+\Omega_1) \cdot \Omega_0^n \left (1+\frac{\Omega_1}{\Omega_0}\right )^n \cdot (1+\delta(t))
\end{eqnarray} 
Only keeping linear terms of $\delta$ or $\Omega_1$, $k(\Omega_0+\Omega_1) = k(\Omega_0) + \frac{{\rm d} k}{{\rm d} \Omega}|_{\Omega_0} \Omega_1$, $(1+\frac{\Omega_1}{\Omega_0})^n = 1+n \frac{\Omega_1}{\Omega_0}$. After some recollection, the equation for $\dot{\Omega}_1$ is
\begin{equation}\label{Omegadot_final}
\dot{\Omega}_1 + \beta \Omega_1 = A(t),
\end{equation}
where $\beta = (\frac{{\rm d} k}{{\rm d} \Omega} |_{\Omega_0} + k(\Omega_0) n \frac{1}{\Omega_0}) \Omega_0^n 
= \frac{n_{\rm steady}}{2} \frac{1}{\tau_{\rm c}}>0$ is the dissipation term, and $A(t) = -k(\Omega_0) \Omega_0^n \delta(t) = \dot{\Omega}_0 \delta(t)$ is the fluctuation term. Equation (\ref{Omegadot_final}) is the same as the Langevin equation for Brownian motion. It may be called ``Langevin equation for pulsar spin-down''. 

From equation (\ref{Omegadot_final}), $\Omega_1$ is a Markov random process. Since there are many degrees of freedom in the magnetosphere which result in the fluctuation term, $\Omega_1$ is a Gaussian random process. Therefore, $\Omega_1$ is a Gaussian, Markov random process. The spectral density for a Gaussian, Markov process has the general form (Thorne \& Blandford 2016)
\begin{equation}
\label{spectral_density_GaussianMarkov}
S_{\Omega_1} (f) = \frac{4\sigma_{\Omega_1}^2/\tau_{\rm r}}{(2\pi f)^2+ (1/\tau_{\rm r}^2)},
\end{equation}
where $\sigma_{\Omega_1}^2$ is the variance of $\Omega_1$, and $\tau_{\rm r}$ is the relaxation time. In the present case, the relaxation time is approximately the dissipation time scale $\tau_{\rm r} \sim \tau_{\rm c}$. This will be proved later by solving equation (\ref{Omegadot_final}) directly. The observational time scale of pulsars are generally smaller than the relaxation time $f^{-1} \ll \tau_{\rm r}$ or $\tau_{\rm r}^{-1} \ll f$. This is contrary to the Brownian motion case, where the observational time scale are much larger than the relaxation time scale. The spectral density for $\Omega_1$ can be simplified 
\begin{equation}
S_{\Omega_1}(f) = \frac{4\sigma_{\Omega_1}^2/\tau_{\rm r}}{(2\pi f)^2}. 
\end{equation}
This approximation is equal to neglecting the dissipation term in equation (\ref{Omegadot_final}) and $\Omega_1$ has a random-walk type spectral density. The spectral density for $\dot{\Omega}_1$ is 
\begin{equation}
\label{spectral_density_Omega1dot}
S_{\dot{\Omega}_1}(f) = (2\pi f)^2 S_{\Omega_1}(f) = 4 \sigma_{\Omega_1}^2/\tau_{\rm r}. 
\end{equation}
The variance of $\dot{\Omega}_1$ is 
\begin{equation}
\label{variance_Omega1dot}
\sigma_{\dot{\Omega}_1}^2 = \int_0^{\infty} S_{\dot{\Omega}_1}(f) {\rm d}f. 
\end{equation}
In principle, the upper bound in the integration is infinity and the variance of $\dot{\Omega}_1$ diverges. However, for a fluctuation with typical variation time scale $T$ , this may set up an upper limit for the corresponding frequency range $f\le f_{\rm upp}=T^{-1}$. Then the variance of $\dot{\Omega}_1$ is 
\begin{equation}
\sigma_{\dot{\Omega}_1}^2 = \int_0^{f_{\rm upp}} S_{\dot{\Omega}_1}(f) {\rm d}f
= (4\sigma_{\Omega_1}^2/\tau_{\rm r}) f_{\rm upp} = \frac{4\sigma_{\Omega_1}^2}{\tau_{\rm r} T}. 
\end{equation}
Similarly, the spectral density and variance of $\ddot{\Omega}_1$ are 
\begin{equation}
S_{\ddot{\Omega}_1}(f) = (2\pi f)^4 S_{\Omega_1}(f) = 16\pi^2 f^2 \sigma_{\Omega_1}^2/\tau_{\rm r},
\end{equation}
\begin{equation}
\sigma_{\ddot{\Omega}_1}^2 = \int_0^{f_{\rm upp}} S_{\ddot{\Omega}_1} {\rm d}f 
= \frac{16\pi^2 \sigma_{\Omega_1}^2}{3 \tau_{\rm r} T^3}.  
\end{equation}
The rms fluctuation of $\ddot{\Omega}_1$ is 
\begin{equation}
\sigma_{\ddot{\Omega}_1} = \sqrt{\frac{16\pi^2 \sigma_{\Omega_1}^2}{3 \tau_{\rm r} T^3}}
=\frac{1}{\sqrt 3} 2\pi  \frac{\sigma_{\dot{\Omega}_1}}{T}. 
\end{equation}
From equation (\ref{Omegadot_final}) and neglecting the dissipation term $\dot{\Omega}_1 = \dot{\Omega}_0 \delta(t)$, therefore 
\begin{equation}
\label{sigmaOmegadot}
\sigma_{\dot{\Omega}_1} = |\dot{\Omega}_0| \delta,
\end{equation}
where $\delta$ is the rms fluctuation of $\delta(t)$. 
Therefore, the rms fluctuation of $\ddot{\Omega}_1$ is 
\begin{equation}
\label{rms_Omega1_2dot}
\sigma_{\ddot{\Omega}_1} = \frac{1}{\sqrt 3} 2\pi \delta \frac{|\dot{\Omega}_0|}{T}. 
\end{equation}
It is consistent with the result when assuming periodic fluctuations, equation (\ref{nu2dot_periodic_case}). 
From equation (\ref{sigmaOmegadot}), $\sigma_{\dot{\Omega}_1}/|\dot{\Omega}_0| \sim \delta$. 
For fluctuation amplitude $\delta \ll 10^{-2}$, their effect on spin-down rate is very small and 
they mainly contribute to $\ddot{\nu}$ which characterise the level of timing noise. 
For fluctuation amplitude $\delta \ge 10^{-2}$, its effect on the spin-down rate is also observable. 
At the same time, the fluctuating magnetosphere will also result in some variation of pulsar pulse profile. 
Therefore, some correlation between the spin-down rate variations and pulse profile variations are expected.
This may corresponds to the observations of Lyne et al. (2010).  

Equation (\ref{Omegadot_final}) can also be solved directly. For the interested frequency range $\frac{1}{\tau_{\rm r}} \ll f \ll T^{-1}$, $\beta$ and $A(t)$ may be considered as constant. The Fourier transform of equation (\ref{Omegadot_final}) is 
\begin{equation}
(-i2\pi f + \beta) \tilde{\Omega}_1 = \tilde{A},
\end{equation}
where $\tilde{\Omega}_1$ and $\tilde{A}$ are the Fourier transform of $\Omega_1(t)$ and $A(t)$, respectively. 
Therefore, 
\begin{equation}
\tilde{\Omega}_1 = \frac{\tilde{A}}{\beta -i 2\pi f}. 
\end{equation}
The spectral density of $\Omega_1$ is 
\begin{equation}
S_{\Omega_1}(f)  = \frac{S_{A}(f)}{(2\pi f)^2 + \beta^2},
\end{equation}
where $S_{A}(f)$ is the spectral density of $A(t)$. Compared with equation (\ref{spectral_density_GaussianMarkov}), the relaxation time scale is $\frac{1}{\tau_{\rm r}} = \beta =\frac{n_{\rm steady}}{2} \frac{1}{\tau_{\rm c}}$. Therefore, $\tau_{\rm r} \sim \tau_{\rm c}$ for a steady state braking index $1 \le n_{\rm steady} \le 3$. From the definition $A(t)=\dot{\Omega}_0 \delta(t)$, the spectral 
density of $A(t)$ is 
\begin{equation}
S_{A}(f) = \dot{\Omega}_0^2 S_{\delta} (f),
\end{equation}
where $S_{\delta}(f)$ is the spectral density of $\delta(t)$. The fluctuation $\delta (t)$ may has a white noise type spectral density. 
Introducing a high frequency cutoff and denote the rms fluctuation of $\delta(t)$ as $\delta$, then the spectral density of $\delta(t)$ is
\begin{equation}
S_{\delta}(f) = \delta^2 T. 
\end{equation}
Therefore, the spectral density of $\Omega_1$ is 
\begin{equation}
S_{\Omega_1}(f) = \frac{\dot{\Omega}_0^2 \delta^2 T}{(2\pi f)^2 + \beta^2}. 
\end{equation}
For the interested time scale $f^{-1} \ll \tau_{\rm c}$ or  $\frac{1}{\tau_{\rm c}} \ll f$, the spectral density of $\Omega_1$ can be simplified
\begin{equation}
S_{\Omega_1}(f) = \frac{\dot{\Omega}_0^2 \delta^2 T}{(2\pi f)^2}. 
\end{equation}
The spectral density and variance of $\dot{\Omega}_1$ and $\ddot{\Omega}_1$ can be obtained straight forward.
The same result is obtained as in the above calculations. 

\section{Discussions}

Based on the fact that pulsar timing is correlated with changes in pulse shape (Lyne et al. 2010; Keith et al. 2013; Brook et al. 2015),
the fluctuating magnetosphere are used to explain the following two aspects. Firstly, the measured braking indices of eight pulsars are
less than three. Secondly, the abnormal distribution and variety of $\ddot{\nu}$ for $222$ pulsars as well as $15$ magnetars.
Unlike the sudden switch of pulsar spin-down rate (Kramer et al. 2006; Camilo et al. 2012; Lorimer et al. 2012), 
timing noise is a continued process.
It indicates that switched states reflect the sudden increasement of particle wind (Li et al. 2014; Kou et al. 2015)
while timing noise reflects persistent fluctuations of particle wind, which includes periodic
fluctuations (Section 3.2.1) or random fluctuations (Section 3.2.2). 
Therefore, Section 2 and Section 3 are not separated and they are two aspects of one problem.
The values of $\tau_{\rm c}$ and $\delta$ can decide the fluctuations of pulsars, which can be seen in
equation (\ref{Omega2dot_general}). It shows that, for young pulsars with small characteristic age, it will translate to equation (\ref{nwindsteady}).
However, for old pulsars with large characteristic age, it can be written into equation (\ref{nu2dot_periodic_case}). Figure \ref{figPPdot} shows
that these two groups of neutron stars have different characteristic age.

The magnetospheric changes of neutron stars will affect both the pulse profile and spin-down torque. 
The changes of pulse profile may also contribute to the timing noise. 
Therefore, magnetospheric changes alone will contribute two factor to the timing noise.
The magnetospheric changes include the variation of total particle number and particle spatial distribution. 
A change of total particle number density will result in a significant change of spin-down rate
(Li et al. 2014; Kou et al. 2015).
If the total number of particles is the same only the spatial distribution of
particles is changed, then the pulse shape may change a lot while flux density and spin-down rate may change 
only very little (e.g., PSR J1602$-$5100; Brook et al. 2015).
As seen in Figure \ref{fignu2dotmodel}, the fluctuation amplitude of magnetars is
larger than that of normal pulsars. It may be due to magnetars are magnetism-powered while pulsars
are rotation-powered. 
The $\Omega_1$ in Equation (\ref{Omegadot_final}) is a Gaussian, Markov random process.
Markov process may explain serval pulsar timing phenomena, such as intermittency, profile change
and switched state (Cordes 2013). 

Based on the analysis of pulsar timing noise, the fluctuation timescale is chose as $T\sim0.1 \,\rm yr$ (Lyne et al. 2010; Keith et al. 2013). 
Quasi periodic fluctuations of magnetosphere may result in both observed
distribution of frequency second derivative and quasi-periodic timing residual
(Liu et al. 2011). 
Several pulsars may have high variation amplitude $\delta$, which will lead to a considerable change of $\dot{\nu}$ (e.g., 
at $1\%$ level). If fluctuations are divided into high state and low state, the observed two state of $\dot{\nu}$ are easily understood. 
However, for variation amplitude less than $1\%$, this change of torque may remain undetected at present. 

There are several other models are suggested to explain the braking index or timing noise of pulsars. 
A fall-back disk formed from supernova material modulates the spin-down of young
pulsars, which would cause the pulsars to loss energy more quickly and a braking index between $2$ and $3$ 
(Alpar, Ankay \& Yzagan 2001). Later, the modified fall-back disk model
was used to explained the low braking index of PSR J1734$-$3333 (Caliskan et al. 2013; Liu et al. 2014; Chen \& Li 2015)
and the Vela pulsar (Ozsukan et al. 2014). However, this model about pulsar spin-down only
try to explain the steady braking indices of pulsars but not including the variable and wide distribution
of $\ddot{\nu}$ of pulsars and magnetars. In addition, whether every supernova explosion can form a fallback is still a question. 
An alternative theory about braking index is magnetic field evolution
(Chen \& Li 2006; Pons, Vigano \& Geppert 2012; Ho 2015). Some works in this direction
also considered the distribution of $\ddot{\nu}$ as well as long term red noise (Zhang \& Xie 2012; Xie, Zhang \& Liao 2015; Yi \& Zhang 2015).
However, the correlations between pulsar timing and pulse profile point to a magnetospheric origin of pulsar spin-down torque.

\section{Conclusions}

A fluctuating neutron star magnetosphere is considered. The braking indices of eight pulsars and anomalous 
frequency second derivatives of pulsars and magnetars can be understood uniformly in this scenario.
If the characteristic age is small, then the measured frequency second derivative gives a meaningful braking index. 
The analysis is consistent with Kou \& Tong (2015).
The rotational evolution of the eight normal pulsars which have meaningful braking
indices ($1 \leq n \leq 3$) are spun-down by both magnetic dipole radiation and particle wind.
In the $P$-$\dot{P}$ diagram, these eight pulsars will evolve to the death valley
but not to the cluster of magnetars (Figure \ref{allsourcesPPdot}). It indicates that the effect of pulsar death is important
for the long term rotational evolution of pulsars. 
If pulsars have braking indices near $3$ (e.g., PSR B1509$-$58 and PSR J1119$-$6127), the characteristic magnetic field will
have similar values with the derived magnetic field. Because in this case, pulsars are dominated by magnetic dipole
radiation. For the same reason, the  characteristic age will also have similar values with the derived age. 
Therefore, for young pulsars their values of characteristic magnetic field and characteristic age still have reference meaning.
The two low braking index pulsars (the Vela pulsar and PSR J1734$-$3333) are good examples to constrain
different acceleration models in the neutron star magnetosphere (Figure \ref {VelaPPdotall} and Figure \ref {J1734PPdot2all}). 
For PSR J1734$-$3333, except VG(CR), SCLF(I) and OG model, other acceleration models are ruled out.
Also, the effect of pulsar death make important influence at present for PSR J1734$-$3333.

For old pulsars with large characteristic age, their frequency second derivative are dominated by the magnetospheric fluctuations. 
In this case, the frequency second derivative is better to identify the spin-down behavior of pulsars instead of the braking index. 
It can explain the observations of 222 pulsars in Hobbs et al. (2010) and 15 magnetars (e.g., Dib \& Kaspi 2014): (1) the abnormal and wide distribution of
pulsars and magnetars $\ddot{\nu}$; the statistically equal values of positive and negative $\ddot{\nu}$ (equation (\ref{periodicnu2dot})); (2) the $\ddot{\nu}$ can change sign for the same source during different time span of observations (equation (\ref{periodicnu2dot})).
Magnetars always have larger fluctuation amplitude than normal pulsars.
The magnetospheric variations may have the form of random fluctuation. The fluctuating component $\dot{\Omega}_{1}$ is a Gauss, Markov process. This case
has similar results with periodic fluctuation, which confirms the calculations of that case. 
These two different cases show the same distribution of $\ddot{\nu}$.
The magnetospheric fluctuation will influence both pulsar timing and radiation, 
which has already been obtained by the observations (Lyne et al. 2010). 
It indicates that the pulsar braking torque originates from neutron star magnetosphere and it has fluctuations.

\section*{Acknowledgments}

The authors would like to thank R. X. Xu for discussions. H. Tong is
supported by West Light Foundation of CAS (LHXZ201201), 973 Program (2015CB857100), Qing Cu Hui of CAS
and NSFC (U1531137).

\label{lastpage}

\end{document}